\documentstyle{EuroPhys}
\def\etal{{\hbox{{\tenit\ et al.\/}\tenrm :\ }}}

\def\And{{\rm and\ }}

\def\stars{\bigskip\centerline{***}\medskip}

\newif\ifboo \boofalse

\def\Review#1{\boofalse{\it #1},}
\def\Name#1{{\sc #1},}
\def\Vol#1{\ifboo Vol. {\bf #1}\else{\bf #1}\fi}
\def\Year#1{\ifboo #1\else(#1)\fi}

\def\Page#1{\ifboo {\rm p. #1}\else{\rm #1}\fi}
\begin{document}
%
%%%   The headers.
%
%%%   These three macros are to have correct headings in your paper.
%%%   You shall omit all the arguments in the two macros `\euro{}{}{}{}'
%%%   `\Date{}' and fill in `\shorttitle{}'. 
%%%   If there is more than one author in the 
%%%   \shorttitle macro, use the macro \etal after first author's name
%%%   to obtain the correct heading.
%
\euro{}{}{}{1998}
\Date{16 January 1998}
\shorttitle{N. SOI\'{C} \etal $\alpha${} - $^{5}$He{} 
 decaying states of $^{9}$Be}
%
%%%  The title, the Author(s) and the affiliation(s)
%
%%%   The title is set in bold (initial word only is capitalized).
%%%   Mathematical expressions and formulas within the title shall be left
%%%   in light face. Initial(s) of the first name(s) are followed by the
%%%   author(s)'s last name(s). If the authors have different affiliations,
%%%   the name must be followed by one or more \inst{number} each referring
%%%   to one of the addresses to appear in the following macro \institute.
%%%   Other items like `Present address' or `email' may be added by putting
%%%   a `\footnote' after the last \inst{number}.
%%%   Begin each address with \inst{number}; the end of an address is \\;
%%%   \\ can also be used to break a line.
%
\title{$\alpha${} - $^{5}$He{} decaying states and 
 the ground state  rotational band of $^{9}$Be }
\author{N. Soi\'{c}\inst{1}, D. Cali\inst{2},  S. Cherubini\inst{2},
     E. Costanzo\inst{2}, M. Lattuada\inst{2},\\
 \mbox{D\raisebox{0.30ex}
{\hspace*{-0.75em}-}\hspace*{ 0.42em}}. Miljani\'{c}\inst{1},
 S. Romano\inst{2}, C. Spitaleri\inst{2}
  \And M. Zadro\inst{1}}
\institute{
     \inst{1} Ru\mbox{d\raisebox{0.75ex}
{\hspace*{-0.32em}-}\hspace*{-0.02em}}er 
Bo\v{s}kovi\'{c} Institute, Zagreb, Croatia
\\
     \inst{2} INFN-Laboratorio Nazionale del Sud and
 Universit\`a di Catania, Catania, Italy \\}
%
%%%    The `\maketitle' macro needs the following macro:    \rec{}{}
%%%    to be left empty.
%
\rec{}{in final form }
%
%%%   Physics Abstracts Classification.
%
%%%   There are two macros: the first one `\pacs{}' makes the PACS 
%%%   environment,the second one `\Pacs{}{}{}' can be used for each
%%%   classification you need.
%%%   To create the subject index of the volume it is important to divide
%%%   the classification numbers into the three different arguments like
%%%   in the following examples 
%
\pacs{
\Pacs{25}{70-z}{Low and intermediate energy heavy-ion reactions}
\Pacs{27}{20+n}{6$\leq A \leq$ 19}
      }
\maketitle
%
%%%   ! Don't forget this command to format the title page of your article!
%
%%%   The Abstract
%
\begin{abstract}
 In a measurement of the 
 $^{9}$Be{}($^{7}$Li{},$\alpha${}$^{7}$Li{})$\alpha{}$n 
 reaction at {\it E$_{{\rm i}}$} = 52 MeV it is unambigously 
 established for the first time that the  $^{9}$Be{} excited states around
 6.5 and 11.3 MeV decay into the $\alpha{}$ + $^{5}$He channel. This
 fact may support previous claims that the 11.3 MeV state is also
 a member of the ground state rotational band.
\end{abstract}
%
%
%%%   Main text
%
%%%   Sectioning
%
%%%   In EuroPhys there is only ``one'' level of sectioning `\section{}'.
%
%Main text begins here.
%
\section{Introduction}
 $^{9}$Be nucleus has one of the largest deformations
 among the stable nuclei.
 Beside the ground state ($\frac{3}{2}^{-}$) only two other members
 ($\frac{5}{2}^{-}$,
 $\frac{7}{2}^{-}$) of the {\it K} = $\frac{3}{2}^{-}$ 
 rotational band are well established.
 However, there is a standing controversy about 
 the fourth member of the band.
 A state from the doublet with the excitation energies between 11 and 12
 MeV has been sometimes mentioned as possible $\frac{9}{2}^{-}$ member
 (see {\it e. g.} \cite{1}). A bump is visible at these excitations in
 energy spectra from the
 $\alpha$ - particle inelastic scattering (strictly isoscalar excitation
  \cite{2,3,4,5}). Peterson \cite{3} even measured the
 angular distribution for the 11.28 MeV state and concluded
 that it should be the expected  $\frac{9}{2}^{-}$ state. 
 However, recent results on electron and proton
 inelastic scattering \cite{6,7} were interpreted in another way: \\
 i) The level, listed in recent compilation  \cite{8} at 6.76 MeV, was
 separated into two components identified as the
 $\frac{7}{2}^{-}$ member
 of the band located at 6.38 MeV and the $\frac{9}{2}^{+}$
 weak coupling state at 6.76 MeV;
 ii) In the case of the 11 MeV states it was claimed that the 
 (p,p$^{'}$) data did not give any and (e,e$^{'}$) data
 only a weak evidence of
 the state at 11.81 MeV. For the other state at 11.28 MeV
 {\it J}$^{\pi}$ = $\frac{9}{2}^{+}$
 was given as the most likely assignment.
 However, from the delayed $\alpha$ - particle spectra
 of the $^{9}$Li{} $\beta$ - decay 
 it was concluded that the 11.28 MeV state was a negative 
 parity state, probably $\frac{3}{2}^{-}$ \cite{9}.

 Recently, the microscopic multicluster model \cite{10} was applied to
 the study of the mirror nuclei, $^{9}$Be{} and $^{9}$B{}. They were
 described in a three - cluster model comprising two 
 $\alpha$ - particles
 and a single nucleon. From this calculation it was concluded that the
 $^{5}$He{} + $\alpha$ configuration is a very good 
 approximation for the
 $^{9}$Be{} ground state wave function. There have also
 been many studies
 on the $\alpha$ - $\alpha$ quasi-free scattering on
 $^{9}$Be{} \cite{11}.
 They all have found large $\alpha$ - particle
 spectroscopic factors
 for the ground state. In the analysis of the data from these
 studies it has been assumed that the knocked out cluster 
 is bound in a 3{\it S} or 2{\it D} state with equal
 occupation probability.
 It was also claimed that some quasi-free reactions 
 on the $^{5}$He{}
 cluster in $^{9}$Be{} were observed, like ($^{3}$He{},t) and
 ($^{3}$He{},$\alpha$) reactions  \cite{12,13}.

 All these results point out that one could expect that the unbound
 states of the rotational band having similar structure to the
 ground state could preferentially decay into the $\alpha$ - 
 $^{5}$He{} channel. In the most recent compilation of $^{9}$Be{}
 states \cite{8} there is no mention of $\alpha$ - decay
 from states
 between 6 and 7 MeV as well as between 11 and 12 MeV.
 From $^{9}$Li{} ($^{9}$C{}) $\beta$ - decay studies \cite{14,15}
 there have been claims 
 that some states in $^{9}$Be{} ($^{9}$B{})
 decay into the $\alpha$ + $^{5}$He{} ($\alpha$ + $^{5}$Li{})
 channel.
 In this way Nyman {\it et. al.}  \cite{14} could  explain the
 high energy part of the $\beta$ - delayed $\alpha$ - particle
 spectra either by the decay of the 11.81 MeV state alone or by
 addition of 30 \% contribution of the 11.28 MeV
 state. This is in contrast to the explanation
 of the same spectra given by Langevin 
 {\it et. al.} \cite{9}, who invoke simultaneous breakup of the 
 11.28 MeV state into 2 $\alpha$ + n. Both papers did
 not explore the third possibility of decay of the
 $^{9}$Be{} 11 MeV state(s):
 the neutron emission into the low energy tail of the 
 very broad  11.4 MeV state of
 $^{8}$Be{}. Because of all this  one  cannot claim unequivocally
 from the $^{9}$Li{} data that $^{9}$Be{} 11 MeV states decay into
 the $\alpha$ + $^{5}$He{} channel. In order to learn
 more about these 
 states and their $\alpha$ - $^{5}$He{} decay one should perform 
 kinematically complete measurements of the processes involving 
 these states.

\section{Experiment}
 Recently, an experiment has been performed to study different $^{9}$Be{}
 + $^{7}$Li{} reactions \cite{16}. In the experiment a 52 MeV $^{7}$Li{}
 beam from the SMP Tandem Van de Graaff accelerator (Laboratorio
 Nazionale del Sud) was used to bombard a self-supported
 beryllium target 
 (400 $\mu$g cm $^{-2}$). The outgoing charged particles
  were detected and identified
 in several particle telescopes consisting either of silicon
 surface barrier ( $\Delta${\it E} and {\it E}) detectors (SDT)
 or of ionization chamber and position sensitive silicon 
 detector (ICPSDT).  Coincidence events between any two
 telescopes were recorded. At
 the beginning of the experiment single telescope spectra were
 recorded.  Only a part of the data is discussed here.

\section{Results and discussion}
 A typical {\it Q} - value spectrum for the 
 $^{9}$Be{}($^{7}$Li{},$^{7}$Li{})$^{9}$Be{} reaction
 is shown on fig. 1. 
 The spectrum was obtained from the $^{7}$Li{} data
 measured with the
 ICPSDT positioned at 24$^{\rm o}$ with an opening angle of
 4$^{\rm o}$. Four
 distinct groups are observed corresponding to
 $^{9}$Be{} excitation
 energies of 0.0, 2.4, 6.4 and 11.3 MeV. The satellites
 clearly seen for
 the first two groups correspond to the simultaneous
 excitation of the
 $^{7}$Li{} projectile to its first excited state (478 keV).
 Similarly to the spectra from $\alpha$ - particle
 scattering on 
 $^{9}$Be{} \cite{2,3,4,5}, the $^{7}$Li{}
 spectra show strong excitation of
 the states with energies approximately following the 
 {\it C} [{\it J}({\it J}+1) - 
 {\it J}$_{\rm 0}$({\it J}$_{\rm 0}$+1)]
 rule ({\it C} is a constant and {\it J}, {\it J}$_{\rm 0}$ the
 spins of excited and ground state, respectively).

 In the analysis of the $\alpha$ - $^{7}$Li{} coincidence
 data one 
 could notice that a large part of the yield from the
 $^{9}$Be{}($^{7}$Li{},$^{7}$Li{}$\alpha$)$\alpha$n reaction  is
 concentrated on the kinematical loci corresponding to the 
 $^{5}$He{} ground state ({\it Q} = -2.5 MeV).
 This is illustrated on fig. 2. by the  {\it Q} - value spectra
 obtained from the  $^{7}$Li{} - $\alpha$
 coincidence data measured at the detectors angle pair
 24$^{\rm o}$ - 50$^{\rm o}$.
 The full line represents the spectrum with
 all the data taken into account and the dashed
 line is the one with only the data having 
 $\alpha$ - particle 
 energies above 7 MeV {\it i.e.} excluding the contribution
 of the strongly
 excited 2.43 MeV state of $^{9}$Be{}, which decays into the n +
 $^{8}$Be{} channel. 
 One can also notice a broad structure around {\it Q} = -6 MeV which
 may be due to the reactions leading to the first excited state
 of $^{5}$He{}. 

  Fig. 3. shows the $^{9}$Be{} excitation energy spectrum obtained
  from the $^{7}$Li{} - $\alpha$ coincidence data, measured with an
 ICPSDT at 26$^{\rm o}$ and a SDT at 50$^{\rm o}$,
 with a cut imposed on the 
  {\it Q} - value spectrum corresponding to the $^{5}$He{}
 ground state.
 The higher yield for {\it E}$_{{\rm x}}$
 between 4 and 15 MeV is in part due
 to the quasi-free $^{7}$Li{} - $\alpha$ scattering contribution. 
 This contribution was also observed for all those angle 
 pairs satisfying 
 kinematical quasi-free scattering condition \cite{16}.
 Two distinct peaks are also 
 visible corresponding to the excitation energies of
 6.5 and 11.3 MeV.
 From the spectra it was not possible
 to conclude, whether the broad peak centered between 6 and
 7 MeV is due
 to two levels at these excitations as claimed in \cite{6,7}
 or to a single broad state \cite{8}.
 It may be added that three different theoretical calculations
 \cite{10,17,18} predict the  $\frac{7}{2}^{-}$ and
 $\frac{9}{2}^{+}$ states at these energies to be very close to
 each other ($\Delta$E $\leq$ 400 keV). In the most recent one
 \cite{10} the widths were also calculated and the
 $\frac{9}{2}^{+}$ state was a factor of 2.4 wider than the 
 $\frac{7}{2}^{-}$ state (2.9 and 1.2 MeV). However, the 
 same theories also predict several other states of 
 $^{9}$Be but their existence has not been confirmed experimentally.
 For illustration it is also shown the result of a calculation
 with only two Breit - Wigner terms (multiplied by the
 phase space factor) with the following excitation energies 
 and widths of these states:
 {\it E}$_{{\rm 1}}^{{\rm exc}}$ = 6.76 MeV 
 $\Gamma _{{\rm 1}}$ = 1.540 MeV 
 {\it E}$_{{\rm 2}}^{{\rm exc}}$ = 11.28 MeV
 $\Gamma _{{\rm 2}}$ = 0.575 MeV as quoted in \cite{8}. 
 The tails of the peaks are probably due to the excitation of 
 $^{9}$Be{} to these states together with the $^{7}$Li{} excitation
 to its first excited state as observed also in the $^{7}$Li{} single
 spectra. 
 Obviously, from these measurements it is not possible
 to say anything about the spins and parities of the
 observed levels.

 The experimental results unambiguously show for
 the first time that
 these states of $^{9}$Be{} decay into
 $^{5}$He$_{{\rm g.s.}}${}
 + $\alpha$, which is also the predominant configuration
 of its ground state.
 This fact and their strong excitation in $\alpha$
 and $^{7}$Li{}
 inelastic scatterings together with their energies
 following the "rotational formula" may support the claims
 that both of
 them belong to the ground state rotational band.
 However, if the structure around 6.5 MeV is due to only one
 state, the third member of the band, then its larger width
 with respect to the one of the expected fourth member at 
 11.3 MeV would seem to contradict these claims.
 Also, in present data there is no evidence for
 excitation of possible $\frac{11}{2}$ member which should fall
 around 17 MeV.
 Obviously, final conclusion on the higher members of the
 band should wait for additional experimental and theoretical
 results.

%
%%%   In the acknowledgments, use the following macro  before and instead 
%%%   of  ``Acknoledgments''
%
\stars

%
%%%   Bibliography environment begins here. You can use the macros \Name{},
%%%   \And, \Book{} or \Review{}, \Vol{}, \Year{} and \Page{}, to type your
%%%   references.
%

\stars

\newpage

\begin{figure}
\vbox to 1cm{\vfill\centerline{\fbox{Here is the figure 1.}}\vfil}
\caption{Q - value spectrum for the 
 $^{9}$Be{}($^{7}$Li{},$^{7}$Li{})$^{9}$Be{}
 scattering measured at E$_{{\rm i}}$ = 52 MeV and 22$^{\rm o} \leq 
 \Theta \leq $ 26$^{\rm o}$ }
\label{fig1}
\end{figure}

\begin{figure}
\vbox to 1cm{\vfill\centerline{\fbox{Here is the figure 2.}}\vfil}
\caption{Q - value spectra for the 
 $^{9}$Be{}($^{7}$Li{},$^{7}$Li{}$\alpha${})$\alpha${}n
 reaction at E$_{{\rm i}}$ = 52 MeV for 22.5$^{\rm o} \leq
 \Theta _{1} \leq $ 31$^{\rm o}$ and  
 $\Theta _{2}$ = 50$^{\rm o}$, $\Delta \Phi$ = 180$^{\rm o}$
  obtained from all the data (full line) and
  from the data with the contributions from 2.43 MeV 
 state of $^{9}$Be{} excluded (dashed line)}
\label{fig2}
\end{figure}

\begin{figure}
\vbox to 1cm{\vfill\centerline{\fbox{Here is the figure 3.}}\vfil}
\caption{$^{9}$Be{} excitation energy spectra obtained
 from the data for the
 $^{9}$Be{}($^{7}$Li{},$^{7}$Li{}$\alpha${})$^{5}$He$_{gs}$
 reaction  measured at 52 MeV and 26$^{\rm o} \leq 
 \Theta _{1} \leq $ 31$^{\rm o}$ and  $\Theta _{2}$
 = 50$^{\rm o}$, $\Delta \Phi$ = 180$^{\rm o}$.
 The curve represents a calculation with two Breit - Wigner terms. }
\label{fig3}
\end{figure}


\vskip-12pt
\begin{thebibliography}{99}
%
\bibitem{1}
\Name{F. C. Barker } \Review{Nucl. Phys.} \Vol{83}
\Year{1966} \Page{418}.

\bibitem{2}
\Name{R. G. Summers-Gill } \Review{Phys. Rev.} \Vol{109}
\Year{1958} \Page{1591}.

\bibitem{3}
\Name{R. J. Peterson} \Review{Nucl. Phys.} \Vol{A377}
\Year{1982} \Page{41}.

\bibitem{4}
\Name{M. N. Harakeh, J. van Popta, A. Saha
 \And R. H. Siemssen} \Review{Nucl. Phys.} \Vol{A 344}
\Year{1980} \Page{15}.

\bibitem{5}
\Name{Subinit Roy, J. M. Chatterjee, H. Majumdar, S.K. Datta,
 S. R. Banerjee  \And S. N. Chintalapudi} \Review{Phys. Rev. C} 
\Vol{52} \Year{1995} \Page{1524}.

\bibitem{6}
\Name{J. P. Glickman, W. Bertozzi, T. N. Buti, S. Dixit,
 F. W. Hersman, C. E. Hyde-Wright, M. V. Hynes, R. W. Lourie,
 B. E. Norum, J. J. Kelly, B. L. Berman \And D. J. Millener }
 \Review{Phys. Rev. C} \Vol{43}
\Year{1991} \Page{1740}.

\bibitem{7}
\Name{S. Dixit, W. Bertozzi, T. N. Buti, J. M. Finn,
  F. W. Hersman,  C. E. Hyde-Wright, M. V. Hynes, 
 M. A. Kovash, B. E. Norum,  J. J. Kelly, A. D. Bacher,
 G. T. Emery, C. C. Foster, W. P. Jones, D. W. Miller,
 B. L. Berman \And D. J. Millener } \Review{Phys. Rev. C}
 \Vol{43} \Year{1991} \Page{1758}.

\bibitem{8}
\Name{F. Ajzenberg-Selove} \Review{Nucl. Phys.} \Vol{A 490}
 \Year{1988} \Page{1}.

\bibitem{9}
\Name{M. Langevin, C Detraz, D. Guillemaud, F. Naulin,
 M. Epherre, R. Klapisch, S. K. T. Mark, M. de Saint Simon,
 C. Thibault  \And F. Touchard} \Review{Nucl. Phys.} \Vol{A366}
 \Year{1981} \Page{449}.

\bibitem{10}
\Name{K. Arai, Y. Ogawa, Y. Suzuki \And K. Varga} 
\Review{Phys. Rev. C} \Vol{54} \Year{1996} \Page{132}.

\bibitem{11}
\Name{A. A. Cowley, G. F. Steyn, S. V. F\"{o}rtsch,
 J. J. Lawrie, J. V. Pilcher, F. D. Smit  \And D. M. Whittal} 
\Review{Phys. Rev. C} \Vol{50} \Year{1994} \Page{2449}
 and references therein.

\bibitem{12}
\Name{K. Kadija, G. Pai\'{c}, B. Antolkovi\'{c},
 A. Djaloeis \And J. Bojowald} \Review{Phys. Rev. C} \Vol{36}
 \Year{1987} \Page{1269}.

\bibitem{13}
\Name{ M. Lattuada, F.Riggi, C. Spitaleri, 
 D. Vinciguerra, 
 \mbox{D\raisebox{0.30ex}{\hspace*{-0.75em}-}\hspace*{ 0.42em}}. 
 Miljani\'{c}, M. Zadro \And J. Yao}
 \Review{Nucl. Phys.} \Vol{A 458} \Year{1986} \Page{493}.

\bibitem{14}
\Name{G. Nyman, R. E. Azuma, P. G. Hansen, B. Jonson,
 P. O. Larsson, S. Matisson, A. Richter, K. Riisager, O. Tengblad,
 K. Wilhelmsen  \And the ISOLDE Collaboration} \Review{Nucl. Phys.}
 \Vol{A 510} \Year{1990} \Page{189}.

\bibitem{15}
\Name{D. Mikolas, B. A. Brown, W. Benenson, L. H. Harwood,
 E. Kashy, J. A. Nolen Jr., B. Sherrill, J. Stevenson, 
 J. S. Winfield, Z. Q. Xie  \And R. Sherr} \Review{Phys. Rev. C}
 \Vol{37} \Year{1988} \Page{766}.

\bibitem{16}
\Name{N. Soi\'{c}, S. Cherubini, E. Costanzo, M. Lattuada,
 \mbox{D\raisebox{0.30ex}{\hspace*{-0.75em}-}\hspace*{ 0.42em}}. 
 Miljani\'{c}, S. Romano, C. Spitaleri \And M. Zadro}
 to be published.

\bibitem{17}
\Name{H. Furutani, H. Kanada, T. Kaneko, S. Nagata, H. Nishioka,
 S. Okabe, S. Saito, T. Sakuda and M. Seya} \Review{Prog. Theor.
 Phys. Suppl.} \Vol{68} \Year{1980} \Page{193}.


\bibitem{18}
\Name{V. T. Voronchev, V. I. Kukulin, V.N. Pomerantsev, Kh. D. 
 Razikov and G. G. Ryzhikh} \Review{Yad. Fiz.}
 \Vol{57} \Year{1994} \Page{1964}.

\end{thebibliography}
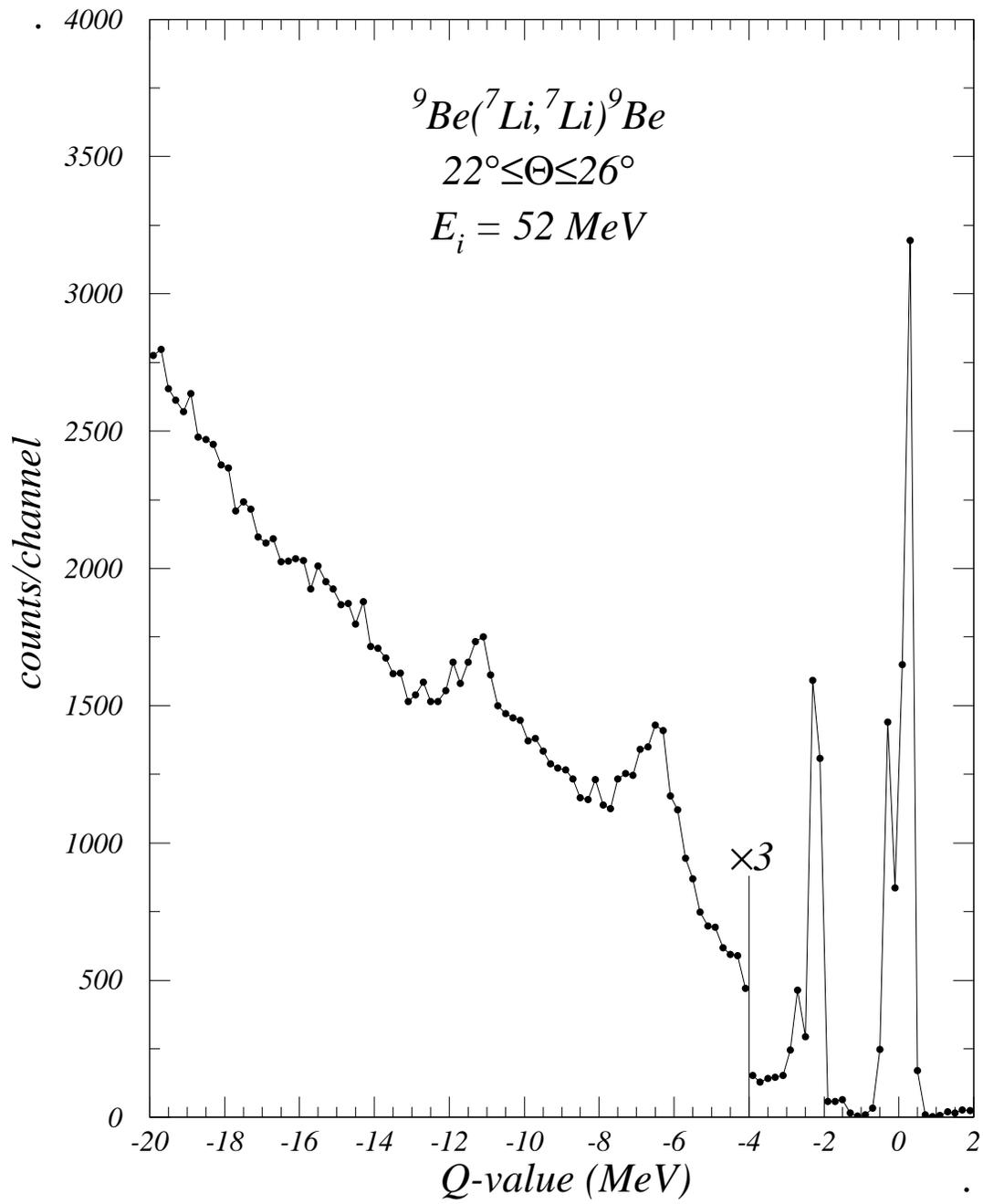
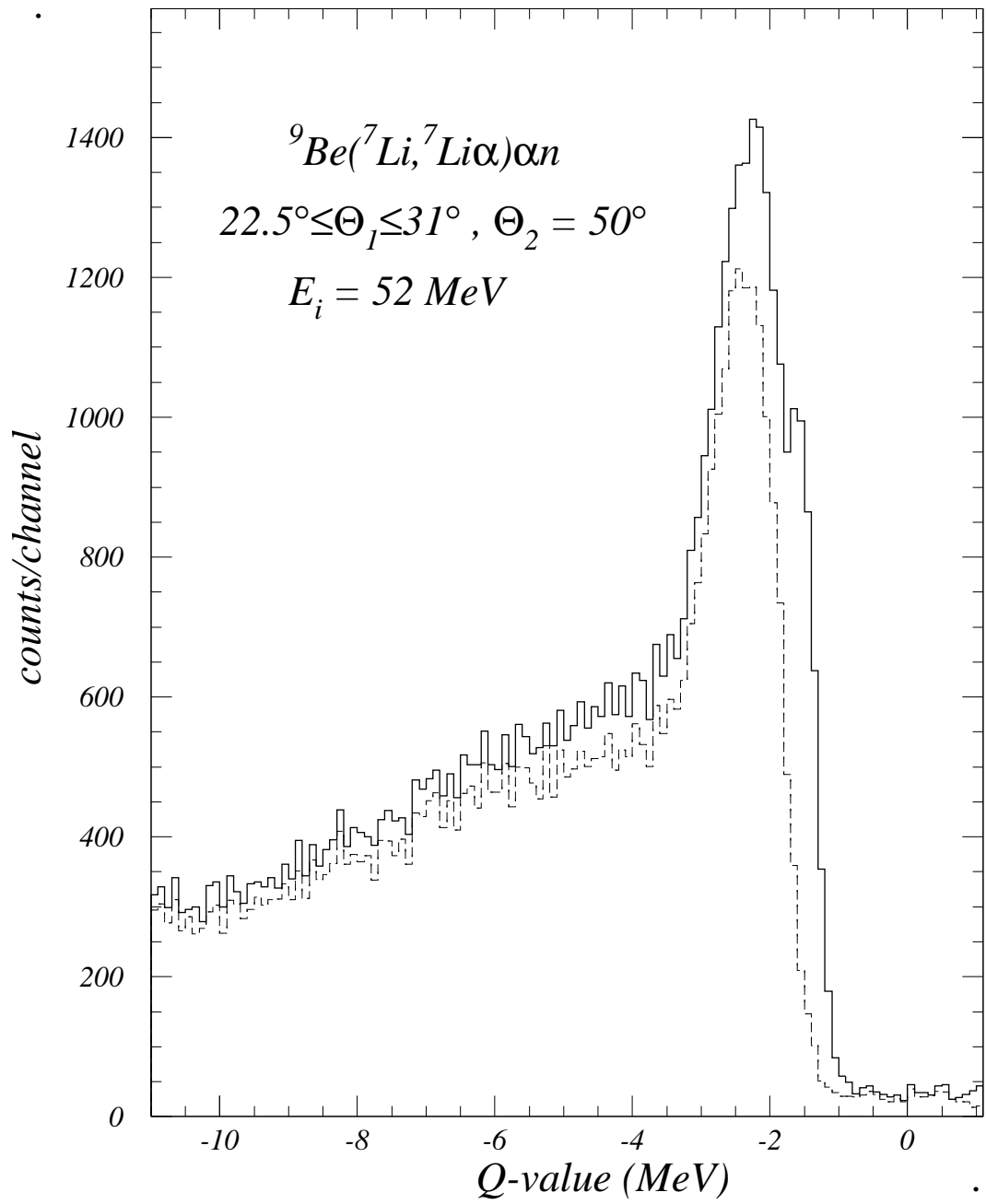
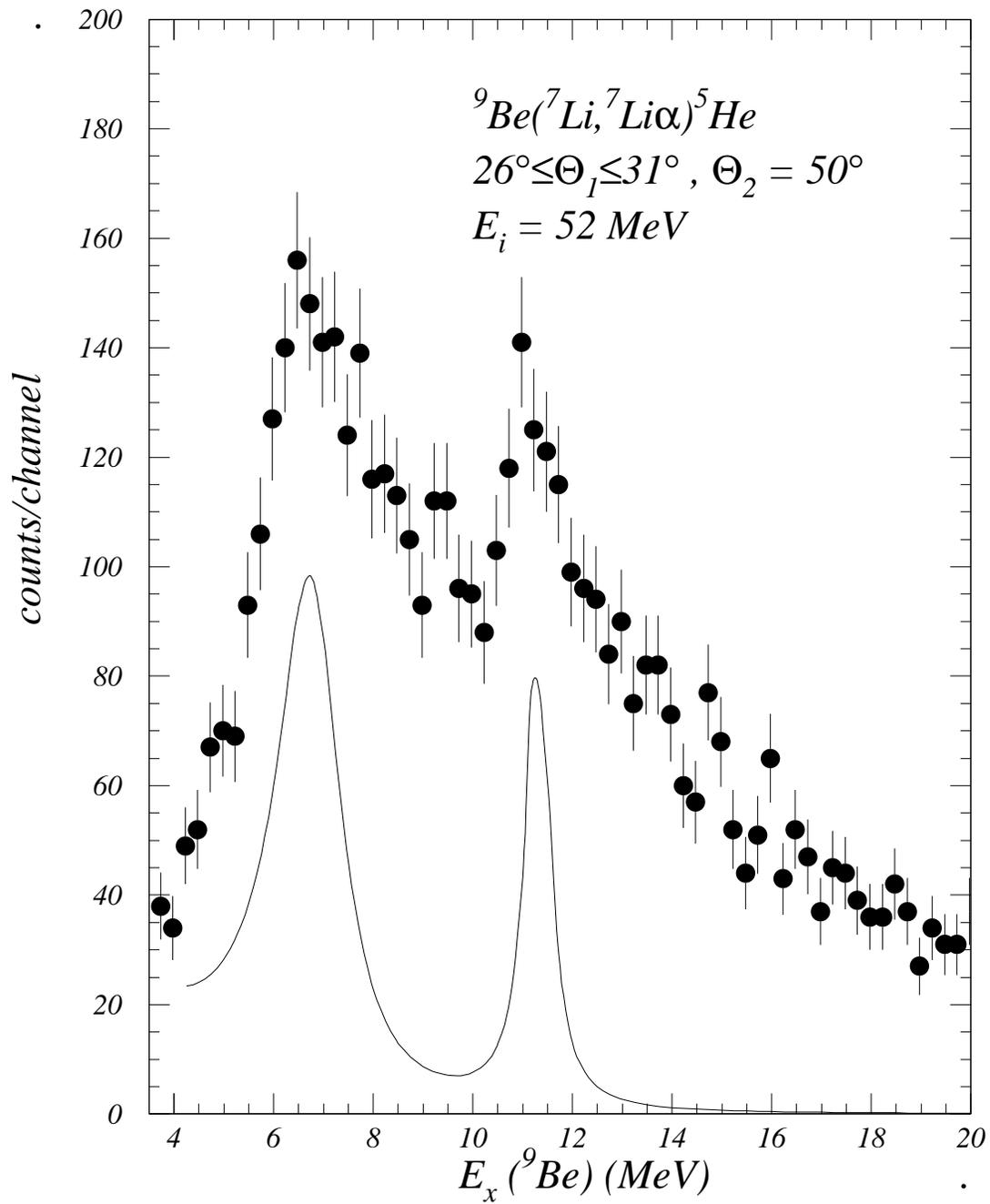
\end{document}